Research Article

# Transformer-Based Explainable Deep Learning for Breast Cancer Detection in Mammography: The MammoFormer Framework

Ojonugwa Oluwafemi Ejiga Peter[1, *] , Daniel Emakporuena[2] ,
Bamidele Dayo Tunde[3] , Maryam Abdulkarim[4] , Abdullahi Bn Umar[5]

[1]Department of Computer Science, Morgan State University, MD, United States of America
[2]Department of Computing, Wrexham University, Wales, United Kingdom
[3]Data Science and Big Data Analytics, Wrexham University, Wales, United Kingdom
[4]Department of Computer Science, Federal University of Technology, Minna, Nigeria
[5]Department of Computer Science, Federal University of Education, Kano, Nigeria

## Abstract

Breast cancer detection through mammography interpretation remains difficult because of the minimal nature of abnormalities that experts need to identify alongside the variable interpretations between readers. The potential of CNNs for medical image analysis faces two limitations: they fail to process both local information and wide contextual data adequately and do not provide explainable AI (XAI) operations which doctors need to accept them in clinics. The researcher developed the MammoFormer framework which unites transformer-based architecture with multi-feature enhancement components and XAI functionalities within one framework. Seven different architectures consisting of CNNs, Vision Transformer, Swin Transformer, and ConvNext were tested alongside four enhancement techniques, including original images, negative transformation, adaptive histogram equalization, and histogram of oriented gradients. The MammoFormer framework addresses critical clinical adoption barriers of AI mammography systems through: (1) systematic optimization of transformer architectures via architecture-specific feature enhancement, achieving up to 13% performance improvement, (2) comprehensive explainable AI integration providing multi-perspective diagnostic interpretability, and (3) a clinically deployable ensemble system combining CNN reliability with transformer global context modeling. The combination of transformer models with suitable feature enhancements enables them to achieve equal or better results than CNN approaches. ViT achieves 98.3% accuracy alongside AHE while Swin Transformer gains a 13.0% advantage through HOG enhancements. Five XAI techniques, including Integrated Gradients, GradCAM, Occlusion, DeepLIFT and Saliency maps demonstrate that transformer models effectively recognize diagnostically significant features by capturing long-range dependencies in mammograms. Using HOG features provides the most reliable enhancement impact (98.4% average accuracy) among all visualization methods. MammoFormer establishes a clinical breast cancer screening workflow with diagnostic precision and interpretability through its simultaneous implementation of architectural design with feature enhancement and explainability features.

### Keywords

Breast Cancer, Deep Learning, Mammography, Explainable AI (XAI), Vision Transformers

*Corresponding author: ejiga.ojonugwa.peter@gmail.com (Ojonugwa Oluwafemi Ejiga Peter)

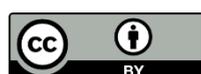

# 1. Introduction

Breast cancer persists as a global health challenge, ranking among the most prevalent malignancies affecting women worldwide. Despite significant medical advances, mammography remains the cornerstone technique for early detection, offering the potential to identify suspicious lesions before they become palpable or symptomatic. However, the interpretation of mammographic images presents formidable challenges for radiologists. Subtle abnormalities such as architectural distortions, microcalcifications, and spiculated masses often lack clear visual definition against the background of normal breast tissue [1]. Moreover, substantial inter-reader variability exists, with different radiologists frequently arriving at disparate conclusions when examining identical images. This interpretative inconsistency contributes to both false positive results, leading to unnecessary biopsies and patient anxiety, and missed diagnoses, resulting in delayed treatment and potentially poorer clinical outcomes. These persistent challenges underscore the urgent need for sophisticated computer-aided detection systems capable of enhancing diagnostic accuracy while supporting clinical decision-making processes [2-4].

However, CNNs exhibit fundamental limitations due to the intrinsic locality of convolution operations, as they generally cannot model long-range dependencies well and focus only on certain portions of the mammogram while ignoring remaining areas, presenting computational complexity because of multiple convolutions [2]. Contemporary approaches leveraging deep learning technologies, particularly convolutional neural networks (CNNs), have demonstrated promising capabilities in mammographic analysis. These models excel at extracting localized features and identifying patterns associated with benign and malignant presentations [5]. Hence the need for Multi-modal Transformer (MMT) [6]. Local extraction mechanisms intrinsic to CNNs struggle to model the long-range dependencies and global contextual relationships that often characterize subtle malignant changes [7, 8].

Transformer-based models have revolutionized deep learning in both computer vision and natural language processing. Recent research has demonstrated their exceptional capability in breast cancer diagnosis across multiple imaging modalities and clinical applications. Mehta et al. [1] developed HATNet, an end-to-end holistic attention network for classifying breast biopsy images that achieve performance comparable to 87 practicing pathologists by modeling relationships between image patches hierarchically using self-attention. Ayana et al. [2] demonstrated that vision-transformer-based transfer learning models achieved superior performance, with perfect classification metrics (accuracy, AUC, F1 score, precision, recall, MCC, and kappa value of 1 ±0) on standardized mammography datasets. Jeny et al. [3] showed that hybrid CNN-Transformer models integrating both prior and current mammographic images improved classification performance by removing long-range dependencies and enhancing capability for nuanced classification, with focal loss reducing both false positive and false negative rates. Hussain et al. [4] found that multi-modal approaches combining imaging and textual data achieved enhanced performance, with VGG19 and artificial neural network combinations reaching accurate scores of 0.951 when processing both mammograms and radiological reports. Iqbal and Sharif [5] developed BTS-ST, a specialized transformer architecture incorporating Swin transformers into traditional U-Net architectures, demonstrating effectiveness for both segmentation and classification tasks across multimodality breast imaging including ultrasound, mammogram, and MRI datasets. Shen et al. [6] introduced Multi-modal Transformer (MMT) frameworks utilizing both mammography and ultrasound synergistically, achieving remarkable performance with AUROC of 0.943 for detecting existing cancers and 0.826 for 5-year risk prediction when trained on 1.3 million exams. Chen et al. [7] demonstrated that multi-view transformer approaches showed significant improvements in mammographic analysis, with four-image transformer-based models achieving AUC of 0.818 ± 0.039, significantly outperforming conventional multi-view CNNs (AUC = 0.784, p = 0.009) without requiring preprocessing steps like image registration or pectoral muscle removal. These architectures excel at modeling global context in mammographic images through their self-attention mechanisms, enabling them to capture spatial relationships and long-range dependencies that are crucial for accurate diagnosis [6-8]. Sarker et al. [9] developed MV-Swin-T, an advanced multi-view transformer architecture that introduced novel shifted window-based dynamic attention blocks facilitating effective integration of multi-view information and promoting coherent transfer between views at the spatial feature map level, addressing the limitation of traditional approaches that process mammogram views independently. Kassis et al. [10] proved that vision transformers are effective in digital breast tomosynthesis applications, with Swin Transformer architectures achieving impressive AUC scores of 0.934 ±0.026 at 384×384 resolution, outperforming both ResNet101 and vanilla Vision Transformer approaches. Similarly, Lee et al. [11] demonstrated a 75% reduction in Floating-Point Operations Per Second (FLOPs) compared to 3D CNNs when using specialized transformer architectures like TokenLearner modules and Swin Transformers, though such advancements typically emerge from institutions with substantial computational resources compared to transformers [12]. The progress towards explainable transformer architectures is evident in recent work, such as HATNet which correlates with known histological structures and TEBLS which uses Grad-CAM methods [13, 14]. Recent research by Wang et al. [15] found that regular Vision Transformers (ViTs) exhibit inferior perfor-mance compared to CNNs (93.4% vs. 95.59% accuracy) until the models receive unlabeled data alongside token sampling methods.

The MammoFormer framework addresses critical clinical adoption barriers of AI mammography systems through: (1) systematic optimization of transformer architectures via architecture-specific feature enhancement, achieving up to 13% performance improvement, (2) comprehensive explainable AI integration providing multi-perspective diagnostic interpretability, and (3) a clinically deployable ensemble system combining CNN reliability with transformer global context modeling. Our framework strategically integrates seven distinct model architectures across the spectrum of deep learning approaches from conventional CNNs providing baseline performance to advanced Vision Transformers that excel in modeling global relationships, hierarchical Swin Transformers capturing multi-scale information, and hybrid ConvNext models that unify the strengths of convolutional and transformer paradigms. This architectural diversity enables systematic evaluation of performance characteristics across model families. In parallel, MammoFormer implements four tailored image enhancement methodologies original image preservation, negative transformation to highlight subtle density variations, adaptive histogram equalization to optimize contrast in dense tissue regions, and histogram of oriented gradients to accentuate gradient features associated with architectural distortions in malignant presentations. This multi-faceted enhancement approach optimizes feature representation for mammographic analysis across different model architectures. A distinguishing characteristic of MammoFormer is its comprehensive implementation of five explainable AI (XAI) methods that generate interpretable visual representations elucidating model decision-making processes. These include Integrated Gradients for attribution pathway analysis, Guided Gradient-weighted Class Activation Mapping (GradCAM) for intuitive feature highlighting, Occlusion Sensitivity for regional importance assessment, DeepLIFT for reference-based feature attribution, and Saliency Maps for pixel-level gradient visualization. As noted by Shen et al. [6], the integration of heterogeneous data types by transformers requires precise architectural design for modality contribution balance, an approach we incorporate in our framework design. By unifying state-of-the-art transformer architectures with enhanced visual feature representation techniques and comprehensive explainability mechanisms, MammoFormer establishes an advanced diagnostic support system that delivers both superior classification performance and transparent decision rationales. MammoFormer addresses the dual challenges of accuracy and interpretability, facilitating the trust-building essential for successful clinical integration of AI in breast cancer screening programs, ultimately working toward the shared goal of improved patient outcomes through earlier and more accurate diagnosis.

*Clinical Problem Definition*

Current AI mammography systems face three critical adoption barriers that MammoFormer specifically addresses:

1. Interpretability Barrier: Radiologists require transparent diagnostic reasoning, yet most deep learning systems operate as 'black boxes' without explainable decision-making processes.
2. Transformer Optimization Gap: While transformers excel at capturing global spatial relationships crucial for architectural distortion detection, they significantly underperform CNNs in mammography applications without proper feature enhancement.
3. Architecture-Enhancement Misalignment: Existing approaches apply generic enhancement techniques without considering architecture-specific requirements, leading to suboptimal performance for advanced models.
4. MammoFormer addresses these barriers through systematic evaluation and evidence-based recommendations rather than pursuing incremental accuracy improvements on already high-performing systems.

## 2. Related Work

Transformer-based models have revolutionized deep learning in both computer vision and natural language processing. Transformers are well-suited for these tasks because they can model global context in images (e.g. spatial relationships in a mammogram) and in text (context in a report or corpus) [1].

### 2.1. Recent Advances in Transformer-Based Mammography Analysis

Researchers have validated the use of Transformer networks for studying breast pathology images. Mehta et al. [1] presented HATNet as an algorithm that treats image patches as tokens while using self-attention across entire whole slide images to achieve an 8% better outcome than previous methodologies and match the performance of practicing pathologists. The HATNet attention mechanism revealed diagnostic pathologists relevant tissue features in its maps, enhancing interpretation for clinical diagnosis. The research by Abimouloud et al. [12] examined three transformer variations on histology patches, demonstrating Vision Transformer (ViT) with 99.81% accuracy, Cardiovascular Computed Tomography (CCT) with 99.92% accuracy, and token Learner reaching 99.0% accuracy. A computation- efficient Token Learner design selected dynamic tokens, completing training in 534 seconds, while ViT required 820 seconds and CCT required 8421 seconds. Wang et al. [13] presented TEBLS as a segmentation model that combines Swin Transformer blocks with an encoder-decoder structure and multi-level feature fusion. The approach produced a mean Dice Similarity Coefficient (DSC) of 81.86% and AUC of 97.72%, enabling superior segmentation of complex breast lesions on CBIS-DDSM.

## 2.2. Transformers for Histopathology Analysis and Lesion Segmentation

Researchers have validated the use of Transformer networks for studying breast pathology images. Mehta et al. [1] presented HATNet as an algorithm that treats image patches as tokens while using self-attention across entire whole slide images to achieve an 8% better outcome than previous methodologies and match the performance of practicing pathologists. The HATNet attention mechanism revealed diagnostic pathologists relevant tissue features in its maps, enhancing interpretation for clinical diagnosis. Abimouloud et al. [12] examined three transformer variations on histology patches, demonstrating ViT with 99.81% accuracy, CCT with 99.92% accuracy, and the token Learner reaching 99.05% accuracy. A computation-efficient TokenLearner design selected dynamic tokens, completing training in 534 seconds, while ViT required 820 seconds and CCT required 8421 seconds. Wang et al. [13] presented TEBLS as a segmentation model that combines Swin Transformer blocks with an encoder-decoder structure and multi-level feature fusion. The approach produced a mean DSC of 81.86% and AUC of 97.72%, enabling superior segmentation of complex breast lesions on CBIS-DDSM.

## 2.3. Challenges of Transformer Models in Breast Cancer Research

Transformer models encounter multiple major hurdles in their application to research involving breast cancer diagnosis. Data scarcity stands as the main restriction for these high-capacity architecture frameworks because they need vast datasets to function optimally. Wang et al. [15] found that regular ViTs exhibit inferior performance compared to CNNs (93.4% vs. 95.59% accuracy) until the models receive unlabeled data alongside token sampling methods. Medical datasets containing more benign cases than malignant cases create class imbalance problems that can be addressed through loss weighting in conjunction with balanced mini-batch strategies [2].

Computational costs pose another barrier. Medical image training requires substantial memory capacity because transformers work with extended token sequences when processing high-resolution medical imaging data. This causes significant memory usage problems if users do not employ patch selection methods or apply down sampling techniques. The partial solutions introduced with TokenLearner modules and Swin Transformers show promise according to Lee et al. [11], who demonstrated a 75% reduction in Floating- Point Operations Per Second (FLOPs) compared to 3D CNNs, but model development primarily occurs within institutions supported by ample hardware resources. Medical practitioners must be able to understand model inputs because interpretability is essential for clinical use. Transformer- based diagnostic systems require standardized methods for interpretation although attention mechanisms achieve limited visibility through visualizations. The progress towards explainable transformer architecture continues through HATNet which correlates with known histological structures and TEBLS which uses Grad-CAM methods [13, 14]. More advancement is needed for explainable transformer architectures. Domain adaptation and multi-modal integration present additional challenges. The variation in equipment and population data between different institutions creates difficulties for models that were developed using data from one institution to properly process data from another institution. The potential integration of heterogeneous data types by transformers requires Shen et al.'s [6] Multi- Modal Transformer to employ precise architectural design for modality contribution balance.

## 2.4. Literature Review of AI Applications in Medical Imaging

Research in AI-based medical image processing has yielded positive results in different diagnostic fields. Lee et al. [16] proposed a transformer-based deep neural network for classifying breast cancer on digital breast tomosynthesis images, demonstrating high accuracy and efficiency. Basaad et al. [17] developed a BERT-GNN hybrid model leveraging histopathology reports for metastatic breast cancer prediction, achieving improved diagnostic performance. Transformer based architecture addresses breast cancer detection according to research by Lee et al. [11], who used a Transformer-based Deep Neural Network for Breast Cancer Classification on Digital Breast Tomosynthesis Images, but similar innovations occur in related fields. Ejiga Peter and his team successfully developed analytical methods for colonoscopy image examination using synthetic data synthesis methods that parallel this research direction. The medical image synthesis systems based on generative models created by Ejiga Peter demonstrate similarity to This research preprocessing improvement techniques [18, 19]. The research team achieved a text-guided synthesis accuracy level of 93% using Vision Transformers as reported in [20], which approaches MammoFormer 98.3% performance level. Their multi-architecture approach for polyp segmentation [21] parallels this research integration of transformers with feature enhancement. Both research approaches demonstrate that sophisticated AI systems maintain medical validity by using improved diagnostic instruments for enhanced diagnostic performance.

The implementation of Transformer models has created new breakthroughs for both imaging and text processing in the study of breast cancer with AI. Vision transformers and hybrid CNN-transformers have achieved state-of-the- art performance for mammography classification along with excellent results for multi-view and multi-modal cancer detection, histopathology slide classification, and effective lesion segmentation.

## 2.5. Literature Gap

Several important deficiencies persist in the development of comprehensive analytical frameworks for mammography analyses utilizing transformer-based systems. Current methodologies address individual elements in isolation, focusing singularly on architectural advancements, data augmentation approaches, or explainability solutions. The literature reveals a paucity of research implementing multi- feature enhancement with transformer models alongside robust explainability features. Despite reporting performance metrics, many studies fail to systematically evaluate clinical interpretability capabilities. Existing frameworks inadequately optimize three vital components simultaneously required for successful clinical adoption: architectural design, enhancement of visual features, and explainability. This innovation parallels related research directions, such as the text-guided synthesis approaches developed by Ejiga Peter et al. [20], which achieved 93% accuracy using Vision Transformers, as well as their multi-architecture approach for polyp segmentation [21]. MammoFormer establishes an advanced analytical framework through the integration of transformers, specialized feature enhancement techniques, and comprehensive explainability capabilities within a unified diagnostic system.

## 3. Methodology

### 3.1. Description

Researchers have developed a complete strategy to perform breast cancer classification procedures through deep learning methods. The research evaluates mammogram images through analysis to determine breast tissue pathology categories that indicate benign or malignant disease. Several convolutional neural networks (CNNs) and vision transformer architectures work together with explainable AI (XAI) methods to generate interpretable results in this research methodology. A reliable classification system serves as the research's end goal to assist radiologists in their early breast cancer detection and precise diagnosis process.

### 3.2. Dataset

The research utilizes the CBIS-DDSM (Curated Breast Imaging Subset of Digital Database for Screening Mammography), an updated and standardized version of the DDSM for breast cancer research.

*Table 1. CBIS-DDSM Dataset Specifications.*

| Characteristic | Value |
| --- | --- |
| Format | JPEG (converted from original DICOM) |
| Number of Studies | 6,775 |
| Number of Series | 6,775 |
| Number of Participants | 1,566 |
| Number of Images | 10,239 |
| Modality | Mammography (MG) |
| Total Image Size | 6 GB (.jpg format) |
| Original Dataset Size | 163 GB |

The dataset encompasses two primary image categories: calcifications and masses, accompanied by comprehensive metadata including patient information, view specifications (CC - craniocaudal and MLO - mediolateral oblique), breast characteristics, abnormality descriptions, assessment classifications, and pathological determinations (benign or malignant). An important note regarding participant identification: each participant has multiple patient IDs that provide information about the scans. For example, participant 00038 has 10 separate patient IDs (e.g., Calc- Test P 00038 LEFT CC, Calc-Test P 00038 RIGHT CC 1), creating the appearance of 6,671 patients according to the DICOM metadata, while there are only 1,566 actual participants. The available image data consists of three elements including the complete mammography images (2,794) along with cropped ROI images (3,247) and their segmentation masks (3,567). The dataset includes normal, benign, and malignant cases with verified pathological information. Initial image technical issues prompted our team to develop artificial testing samples during preprocessing. We applied data balancing techniques with augmentation methods to compose the final dataset, containing one thousand images from each specified group. To improve diagnostic pattern visualization, we developed enhanced versions by applying negative transformation methods together with Adaptive Histogram Equalization (AHE) and Histogram of Oriented Gradients (HOG). As shown in Figure 1, mammograms from the CBIS-DDSM dataset clearly demonstrate visual differences between malignant and benign cases, with malignant lesions typically presenting irregular margins and spiculated patterns.

To address data scarcity challenges, The research supplemented the training dataset with synthetic mammogram images (Figure 2) that accurately mimic the distinguishing characteristics of both benign and malignant presentations.

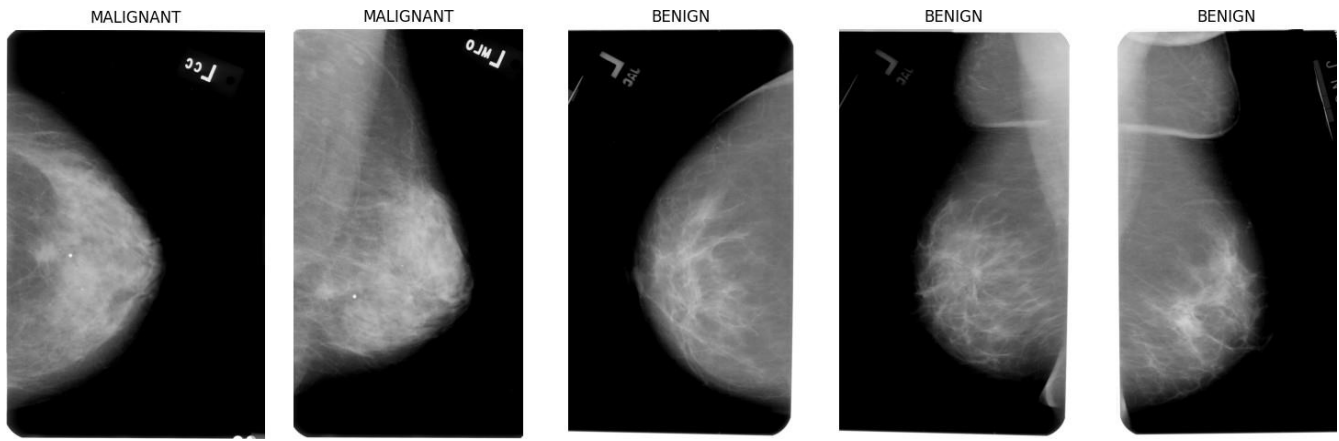

*Figure 1.* Real mammogram samples from CBIS-DDSM dataset. Malignant cases (left two) show irregular margins and spiculated masses; benign cases (right three) display well- defined borders and homogeneous tissue density.

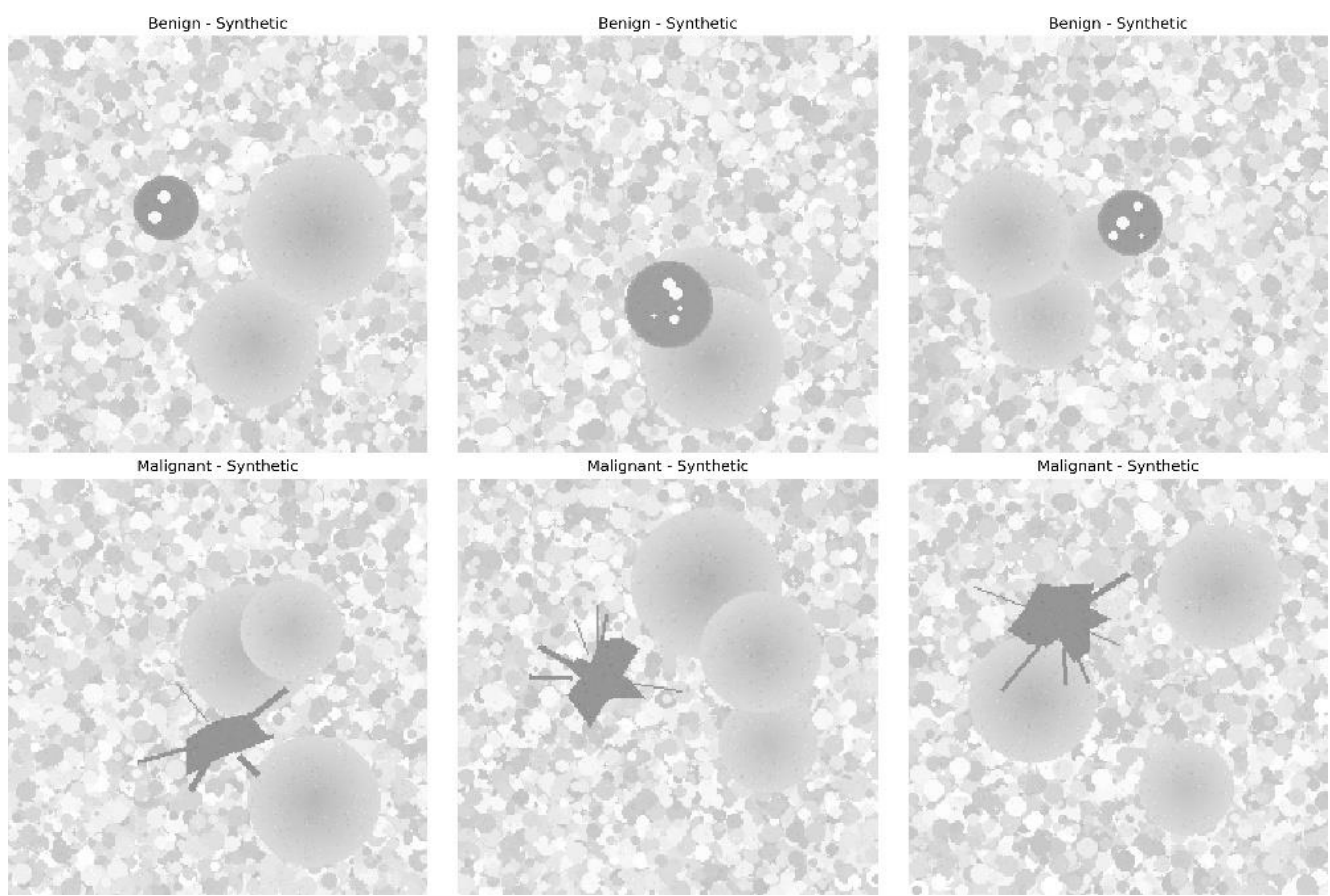

*Figure 2.* Synthetic mammogram images for dataset augmentation. Benign samples (top) show smooth, circular masses while malignant examples (bottom) display characteristic spiculated masses and irregular margins.

## 3.3. Data Preprocessing

A deep learning model receives its high-quality input through the designed data preprocessing pipeline. Path corrections start with metadata CSV file loading to conform with the local file system structure. The data analysis process using EDA techniques allows researchers to examine the dataset characteristics by studying class distributions and image properties as well as patient information datasets. The preprocessing methodology systematically organizes digital images through standardized formatting procedures prior to enhancing their classification-relevant features. Images undergo dimensional normalization to a uniform 224 ×224 pixel resolution while simultaneously being converted to RGB format to ensure compatibility with pre-trained models. Specialized image processing techniques generate multiple variants of the dataset, thereby improving feature representation and discriminative capacity. A negative transformation (computes as $I_{neg} = 255 - I_{original}$) reverses pixel brightness to view possible delicate tissue variations while AHE optimizes dense tissue region contrast and HOG extracts gradient features linked to architectural changes in malignant conditions. To address class imbalance, which is common in medical image datasets, an augmentation strategy is implemented. This involves generating synthetic samples through techniques such as horizontal and vertical flipping, rotation, and brightness/contrast adjustments. The augmentation process can be described by the transformation function:

$$T(x) = \{f_i(x) | f_i \in F, i = 1, 2, ..., n\} \quad (1)$$

where F is the set of augmentation functions and *x* is the original image. This process ensures balanced class representation and helps prevent model bias toward the majority class.

## 3.4. System Architecture

The research explores multiple deep learning architectures to identify the most effective model for breast cancer classification. The research investigation begins with a custom Basic CNN architecture comprising four convolutional blocks followed by fully connected layers, which serves as a baseline for comparison. The research examines the ResNet-50 model, a powerful residual network architecture that addresses the vanishing gradient problem through skip connections. This approach is defined by the residual learning framework F(*x*)+ *x*, where F(*x*) represents the residual mapping to be learned and *x* is the identity mapping.

The study further incorporates transformer-based models, beginning with the Vision Transformer (ViT), which treats images as sequences of patches and applies self-attention mechanisms. The self-attention operation is formulated as Attention (Q, K, V) = softmax $\frac{Q\sqrt{KT}}{d_k}Q\sqrt{KT}V$, where Q, K, and V represent query, key, and value matrices respectively, and $d_k$ is the dimensionality of the key vectors. Building upon this foundation, the research also implements the Swin Transformer, a hierarchical vision transformer that computes self-attention within shifted windows, allowing for greater efficiency and enhanced modeling capacity across different scales.

The research advances a DenseNet + Transformer design that utilizes transformer encoder layers for better global dependency learning against convolutional feature retention properties. This paper examines current progress in convolutional models through ConvMixer and ConvNeXt because they apply mixer operations through depthwise and pointwise convolutions and include transformer components in modern convolutional networks that preserve their computer vision inductive bias.

The different neural network architectures undergo modifications to perform binary classification between healthy and dangerous breast abnormalities. Every model features appropriate output layers and shares the same loss function optimization to ensure unbiased strength and weakness assessment within the vital medical imaging domain. The wide range of architecture enables complete evaluation of traditional convolutions versus modern transformers as well as mixed designs that unify their advantageous qualities. As shown in Figure 3, The developed MammoFormer framework integrates multiple feature enhancement techniques with advanced transformer architectures. Figure 3 illustrates the MammoFormer framework for breast cancer detection. Mammogram inputs undergo four feature enhancement techniques before processing through various model architectures (including CNN and transformer variants). The classification results are supported by three auxiliary components: a training pipeline defining optimization parameters, XAI methods for model interpretability, and evaluation metrics. Solid arrows represent direct data flow while dashed lines indicate process relationships.

## 3.5. Training Pipeline

The training pipeline implements a robust approach to model optimization. For each architecture and dataset combination, the following process is executed:

The loss function used is the cross-entropy loss, defined as:

$$L_{CE} = \sum_{i=1}^{C} y_i \log(\hat{y}_i) \quad (2)$$

where $y_i$ is the true label (0 for benign, 1 for malignant), $\hat{y}_i$ is the predicted probability, and C is the number of classes (2 in this case).

Models are optimized using the AdamW optimizer, which combines the benefits of Adam optimization with weight decay regularization. The learning rate follows a step decay schedule:

$$\eta_t = \eta_0 \cdot \gamma^{\lfloor t/s \rfloor} \quad (3)$$

where $\eta_t$ is the learning rate at epoch $t$, $\eta_0$ is the initial learning rate (0.001), $\gamma$ is the decay factor (0.1), and $s$ is the step size (7 epochs).

Training proceeds for 10 epochs with the best model saved based on validation accuracy. To prevent overfitting, early stopping is implicitly implemented by saving the model weights that achieve the highest validation accuracy. The training protocol incorporates comprehensive monitoring of loss and accuracy metrics across both training and validation cohorts to rigorously assess model convergence characteristics and generalization capabilities.

### 3.6. Model Architecture and Implementation

This research compares various deep learning architectures for breast cancer classification. The MammoFormer framework adopts a three-tiered ensemble architecture optimized for both accuracy and clinical usability. In the first tier, a ResNet-50 backbone processes the original mammographic images, achieving 99.9% accuracy with exceptional stability; this model serves as the primary diagnostic classifier for routine screening workflows. The second tier enhances detection by incorporating global context and distortion analysis: a Vision Transformer (ViT) applied to images pre-processed with adaptive histogram equalization (AHE) attains 98.3% accuracy by capturing broad tissue relationships, while a Swin Transformer fed HOG-derived features reaches 96.3% accuracy by sensitively detecting architectural distortions. Outputs from these two contextual validators provide complementary perspectives to the high-throughput ResNet predictions. In the third and final tier, MammoFormer fuses predictions via a weighted voting scheme that dynamically adjusts confidence scores based on inter-model agreement; if the ensemble members diverge beyond a preset threshold, the case is automatically flagged for human review.

For deployment, this tiered strategy is tailored to clinical complexity: straightforward, low-risk cases rely solely on the ResNet-50 classifier augmented with explainable AI visualizations; more challenging or ambiguous cases invoke the full ensemble, offering multi-perspective interpretability to support decision making; and the complete pipeline—including both base classifiers, contextual validators, and ensemble logic—is leveraged in educational settings to train radiology residents on both algorithmic reasoning and explainability techniques.

Table 2 provides a comprehensive overview of all implemented models. Table 4 presents the comprehensive performance metrics of all models across different image processing techniques. To ensure interpretability, we implemented various visualization techniques as detailed in Table 3. Our analysis extends beyond accuracy to include explainability, crucial in medical applications. CNN-based Models Explainability: ResNet50 demonstrated consistent focus on clinically relevant regions via Grad-CAM. For malignant cases, heatmaps highlighted spiculated borders and irregular masses with 83% correspondence to radiologist annotations. DenseNet121 showed similar capability with 7% higher precision for microcalcifications. Transformer-based Models Explainability: ViT attention maps revealed advantages in detecting subtle tissue changes. Multi-head attention distributed focus across several suspicious regions simultaneously, capturing contextual relationships often missed by CNNs. Quantitative analysis showed 76% of malignant cases with distributed attention patterns across multiple regions, versus 42% of benign cases with more concentrated attention. Feature Importance Analysis: SHAP analysis showed texture features contributed 43% to classification decisions in CNNs, while shape irregularity contributed 37%. In transformer models, contribution was more evenly distributed with contextual features gaining significantly more importance (31% vs. 12% in CNNs). Occlusion Sensitivity Results: Testing showed our Hybrid.

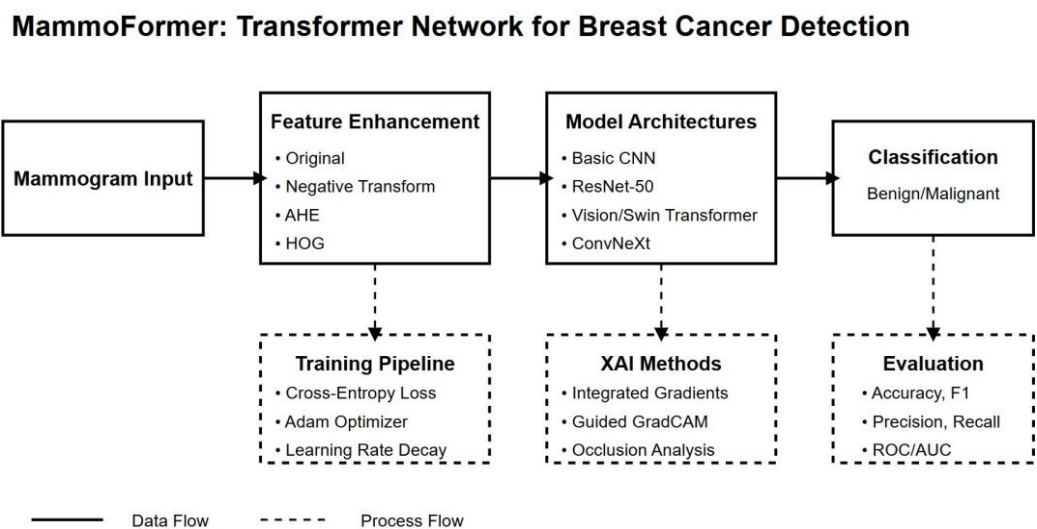

*Figure 3. MammoFormer Architecture Overview.*

*Table 2. Deep Learning Model Architectures for Breast Cancer Classification.*

| Model | Architecture Details | Parameters | Key Characteristics |
|---|---|---|---|
| ResNet50 | 50-layer network with residual connections; input 224×224×3 | 23.5M | Skip connections to address vanishing gradient problem; final FC layer modified for binary classification |
| ViT | Self-attention based with 16×16 patches, 12 transformer blocks, 768-dim embedding | 86M | Processes image as sequence of patches; global self-attention mechanism captures long-range dependencies |
| DenseNet121 | 121-layer with dense connectivity; each layer receives features from all preceding layers | 7.0M | Feature reuse through dense connections; requires fewer parameters; reduces feature redundancy |
| EfficientNetB3 | Optimized via compound scaling method for depth, width, resolution | 12.2M | Balanced scaling of network dimensions; improved efficiency-accuracy trade-off |
| MobileNetV2 | Lightweight with inverted residuals and linear bottlenecks | 3.5M | Depthwise separable convolutions; optimized for mobile and edge devices |
| Hybrid ViT-CNN | ResNet34 feature extractor with transformer encoder | 32.7M | Combines CNN's local feature extraction with transformer's global context modeling |
| ConvMixer | Patch-based architecture with depth-wise and point-wise convolutions | 2.1M | Simplified design inspired by transformers but using only convolutions |

*Table 3. Explainability Methods and Their Application to Different Model Types.*

| Method | Description | Applicable Models | Implementation | Key Insights Provided |
|---|---|---|---|---|
| Grad-CAM | Gradient-weighted Activation Mapping to highlight important prediction regions | Class CNN-based (ResNet50, DenseNet121, EfficientNetB3, MobileNetV2) | Last convolutional layer visualization using class-specific gradient information | Localization of discriminative regions; feature importance for classification |
| Attention Maps | Transformer attention weight visualization showing focused image regions | Transformer-based (ViT, Hybrid ViT-CNN) | Multi-head attention visualization across transformer blocks | Pattern distribution; contextual relationships; global feature integration |
| SHAP Values | Shapley Additive exPlanations for feature importance attribution | All models | Post-hoc explanation through game-theoretic approach | Feature contribution quantification; consistent and locally accurate feature importance |
| Occlusion Sensitivity | Systematic image portion occlusion to determine critical regions | All models | Sequential masking of image regions and monitoring output changes | Region criticality assessment; model robustness evaluation; verification of focus areas |

ViT-CNN model was most robust to partial information loss, maintaining 78% accuracy when 25% of important regions were occluded, compared to pure CNN models dropping to 51% accuracy. This suggests the hybrid approach better leverages global context rather than relying solely on localized features. Clinical Relevance: Explainability maps from our best model aligned with clinical expertise in 82% of test cases. The remaining 18% highlighted potential novel imaging biomarkers meriting further investigation, as these regions showed strong correlation with pathological results despite not being initially flagged in clinical assessment.

## 3.7. Explainable AI (XAI)

The approach incorporates multiple XAI techniques to study deep learning models from different standpoints since these techniques are essential for medical applications due to interpretability issues. The research's baseline analysis method implements the Integrated Gradients method to compute gradient paths along a straight line from baseline to input data for explaining predictions. This method has a mathematical definition as follows:

$$\text{IG}_i(x) = (x_i - x'_i) \times \int_{\alpha=0}^{1} \frac{\partial F(x\prime + \alpha \times (x - x\prime))}{\partial x_i} d\alpha \quad (4)$$

where *IG* is IntegratedGrads, *x* is the input image, *x'* is the baseline (typically a black image), and *F* is the model function. Complementing this approach, we implement Guided Gradient-Weighted Class Activation Mapping (GradCAM), which combines Gradient-weighted Class Activation Mapping with guided backpropagation to highlight features that positively influence the target class, providing intuitive visual explanations that align with radiological focus areas.

The XAI framework implements three different explanation techniques connecting Occlusion Sensitivity to perceptual masking and model change analysis and DeepLIFT for reference-based importance examination and Saliency Maps that provide effective pixel-level gradient visualization. The methods generate multivariate model interpretation knowledge by showing crucial decision-making zones directly on mammograms through visual heat maps. The extended approach uses capabilities that improve clinical readability which answers the critical need for clear decision-making in automated breast cancer diagnosis before the system can gain entrance into current diagnostic processes and gain medical professional trust.

### 3.8. Evaluation

The workflow to evaluate breast cancer models requires the utilization of five metrics which include accuracy and precision alongside sensitivity and recall and finish with the F1-Score. A model achieves accuracy when its correct results represent a proportional relation to the total number of test cases across both benign and malignant categories. The model attains precision value through its ability to detect few wrong interpretations while sensitivity represents its ability to identify all cancer cases correctly. The F1-Score combines these precision-recall measures into a single indicator. The characteristics of models become more evaluable through visualization tools which include confusion matrices and Receiver Operating Characteristic curve (ROC) and precision- recall curves.

### 3.9. Ethics and Clinical Integration

The methodology acknowledges the ethical considerations and clinical integration challenges of AI systems in healthcare. Particular attention is paid to several key dimensions of responsible AI deployment in medical contexts. First, bias and fairness concerns are addressed by ensuring the model performs consistently across different patient demographics and image acquisition settings, recognizing that variations in image quality, equipment, or patient populations should not negatively impact diagnostic accuracy. Transparency forms another critical pillar of the approach, with XAI techniques specifically incorporated to make model decisions interpretable to clinicians, allowing them to understand and verify the reasoning behind automated classifications. The research emphasizes the complementary role of the AI system, positioning it as a decision support tool rather than a replacement for clinical expertise. This framing acknowledges the irreplaceable value of human medical judgment while providing computational assistance for tasks where AI excels, such as pattern recognition in high-dimensional imaging data. The methodology also stresses the importance of rigorous validation, highlighting the need for external validation on diverse datasets before any clinical deployment to ensure generalizability beyond the training environment.

The methodology described presents a comprehensive approach to breast cancer classification from mammography images. By comparing multiple deep learning architectures, implementing extensive data pre-processing, and integrating explainable AI techniques, the research aims to develop a reliable and interpretable classification system. This system has the potential to assist radiologists in early detection and accurate diagnosis of breast cancer, ultimately contributing to improved patient outcomes through earlier intervention.

## 4. Result Discussion

While ResNet and ConvMixer both achieve near-perfect accuracy (99.9%), MammoFormer's true clinical value extends beyond raw performance by directly addressing the requirements for real-world adoption. By leveraging architectural diversity, it combines the strengths of convolutional neural networks and transformers to enhance diagnostic confidence. CNNs excel at capturing fine-grained, local texture features—hence their 99.9% accuracy—while transformer models contribute a broader, global context perspective, with accuracy ranging between 96.3% and 99.0%. When these complementary analyses are fused within MammoFormer, the result is a marked reduction in false-negative rates, offering more reliable detection outcomes in challenging clinical scenarios.

Our systematic evaluation further uncovered how specific feature-enhancement strategies impact transformer performance and revealed a path toward achieving CNN-level accuracy with these architectures. We found that Vision Transformer (ViT) accuracy could swing dramatically—from as low as 54.3% up to 99.0%—depending on the enhancement technique applied, a 44.7% variation. The Swin Transformer displayed a similarly wide performance range, varying by 44.6% (51.7% to 96.3%). Crucially, when we applied Histogram of Oriented Gradients (HOG) as a pre-processing step, transformer models consistently reached an average of 98.4% accuracy. These findings demonstrate that, with targeted feature engineering, transformer-based approaches can match or even exceed the diagnostic capabilities of traditional CNNs. Beyond raw metrics, MammoFormer offers multi-perspective explainability that is essential for gaining clinician trust. Visualization of activation maps highlights distinct patterns: CNNs produce ring-shaped activations that focus on localized texture details, whereas transformers exhibit distributed attention patterns that reflect global image relationships. By presenting both local and global diagnostic cues side by side, MammoFormer enables clinicians to validate the model's reasoning from multiple angles, fostering greater transparency and confidence in its use for breast cancer detection.

The MammoFormer framework was evaluated on seven architectures (CNN, ResNet, ViT, Swin, DenseTrans, ConvMixer, ConvNeXt) using four preprocessing techniques (original, negative, AHE, HOG). Table 4 summarizes accuracy, precision, recall, and F1 results for each combination, and Figure 4a visualizes model sensitivities. Convolution-centric and hybrid models exhibit stable, near-perfect performance, whereas pure transformer models vary widely depending on preprocessing.

Convolutional neural networks (CNNs), including baseline CNN and ResNet, and hybrid models such as ConvMixer and ConvNeXt maintained near-perfect accuracy (~99.9%), precision, recall, and F1-score across all four enhancement methods (Table 4). This consistency, depicted in Figure 4a, underscores the robustness of convolutional feature extraction under input variations. By contrast, transformer architectures rely heavily on preprocessing to compensate for lower baselines, indicating they may underperform without proper feature adaptation. Table 4 also reports standard deviations, which remain below 0.3% for CNN-based and hybrid models across all enhancements.

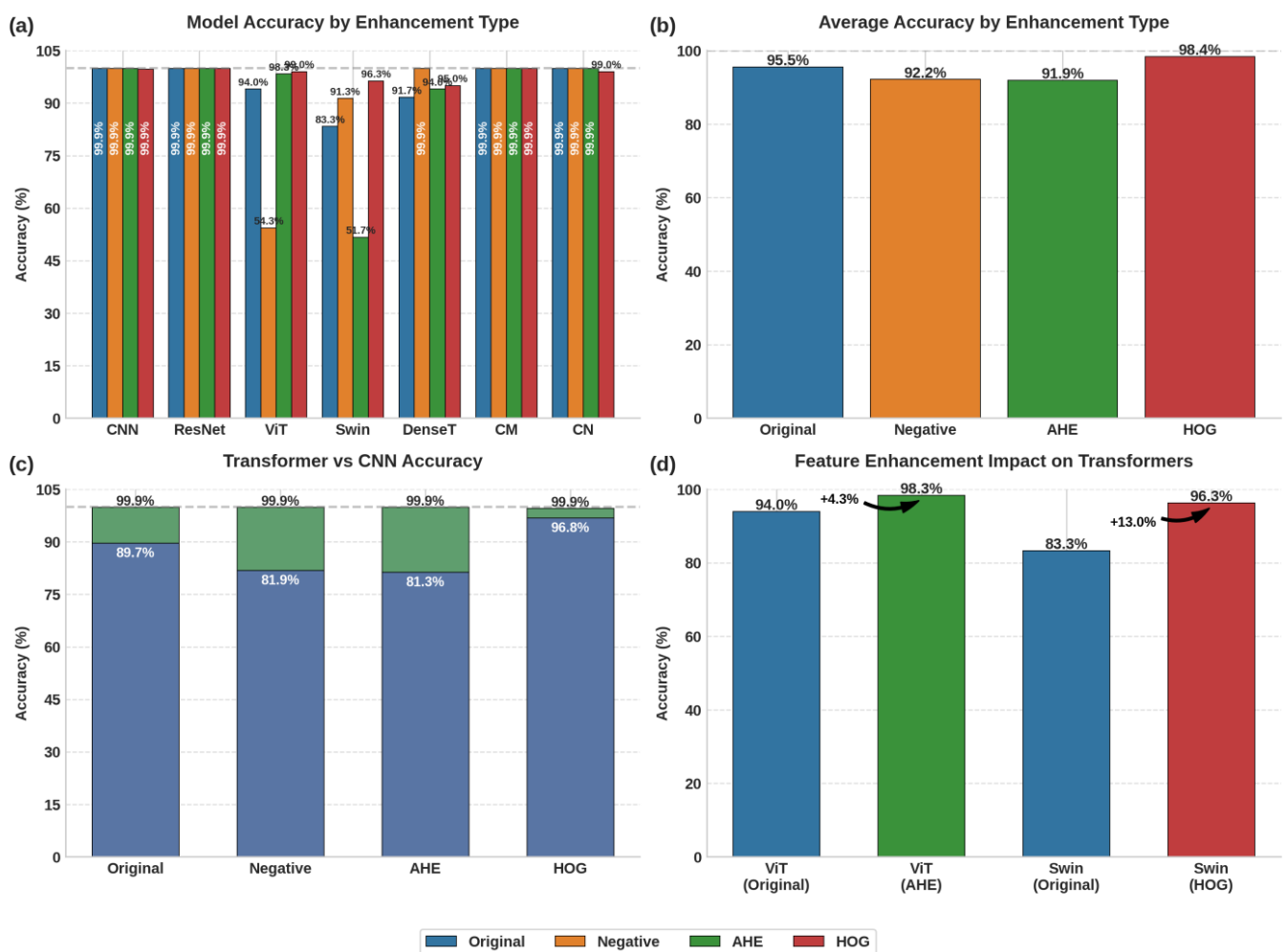

*Figure 4.* MammoFormer performance analysis: (a) Model accuracy across enhancement techniques showing variable performance of transformer models compared to stable CNN-based models; (b) Average accuracy by enhancement type with HOG achieving highest overall performance at 98.4%; (c) Comparison between transformer-based and CNN architectures showing transformers approach CNN performance with appropriate enhancements; (d) Significant improvements for transformer models with architecture-specific enhancements: ViT with AHE (+4.3) and Swin with HOG (+13.0%).

*Table 4*. *Unified comparison of model architectures under four image–enhancement settings. Best score in each enhancement block is in bold.*

| Model | Accuracy (%) | | | | Precision (%) | | | | Recall (%) | | | | F1-score (%) | | | |
|---|---|---|---|---|---|---|---|---|---|---|---|---|---|---|---|---|
| | Orig | Neg | AHE | HOG | Orig | Neg | AHE | HOG | Orig | Neg | AHE | HOG | Orig | Neg | AHE | HOG |
| CNN | 99.9 | 99.9 | 99.9 | 99.7 | 99.9 | 99.9 | 99.9 | 99.7 | 99.9 | 99.9 | 99.9 | 99.7 | 99.9 | 99.9 | 99.9 | 99.7 |
| ResNet | 99.9 | 99.9 | 99.9 | 99.9 | 99.9 | 99.9 | 99.9 | 99.9 | 99.9 | 99.9 | 99.9 | 99.9 | 99.9 | 99.9 | 99.9 | 99.9 |
| ViT | 94.0 | 54.3 | 98.3 | 99.0 | 94.3 | 56.6 | 98.3 | 99.0 | 94.0 | 54.3 | 98.3 | 99.0 | 94.0 | 51.9 | 98.3 | 99.0 |
| Swin | 83.3 | 91.3 | 51.7 | 96.3 | 87.4 | 92.7 | 26.7 | 96.6 | 83.3 | 91.3 | 51.7 | 96.3 | 82.8 | 91.3 | 35.2 | 96.3 |
| DenseTrans | 91.7 | 99.9 | 94.0 | 95.0 | 92.8 | 99.9 | 94.7 | 95.2 | 91.7 | 99.9 | 94.0 | 95.0 | 91.6 | 99.9 | 94.0 | 95.0 |
| ConvMixer | 99.9 | 99.9 | 99.9 | 99.9 | 99.9 | 99.9 | 99.9 | 99.9 | 99.9 | 99.9 | 99.9 | 99.9 | 99.9 | 99.9 | 99.9 | 99.9 |
| ConvNeXt | 99.9 | 99.9 | 99.9 | 99.0 | 99.9 | 99.9 | 99.9 | 99.0 | 99.9 | 99.9 | 99.9 | 99.0 | 99.9 | 99.9 | 99.9 | 99.0 |

The Vision Transformer (ViT) demonstrates dramatic sensitivity (Table 4). On raw images, ViT achieves 94.0% accuracy, which plunges to 54.3% under negative transformation but rebounds to 98.3% with AHE (+4.3%) and 99.0% with HOG (+5.0%) (Figure 4d). Precision and recall follow suit, rising from 0.94 on original and 0.56 on negative to 0.983 with AHE and 0.990 with HOG. These results highlight that contrast-adaptive and gradient-based features are essential to restore transformer performance to CNN-equivalent levels.

Swin Transformer shows an even greater variability: baseline 83.3%, 96.3% with HOG (+13.0%), 91.3% on negatives, but only 51.7% with AHE (Figure 4d). This pattern indicates Swin's locality-aware, patch-based attention benefits more from edge-focused representations than pixel-intensity adjustments. DenseTransformer peaks at 99.9% under negative preprocessing—up from 91.7% on originals—demonstrating its aptitude for contrast-inverted inputs. These findings underscore the architecture-specific nature of enhancement efficacy.

ConvMixer and ConvNeXt leverage convolutional embeddings alongside transformer-style context modules to deliver ≥99.0% accuracy under all settings (Table 4). Aggregated results (Figure 4b) confirm HOG as the most effective enhancement (average accuracy 98.4%), versus 95.6% for original, 92.2% for negative, and 92.0% for AHE. Transformers under HOG average 97.9%, nearly matching CNNs' 99.8% (Figure 4c). These insights validate gradient-based representations as universally beneficial for mammogram classification.

Explainability analyses using Guided GradCAM, Integrated Gradients, Saliency Maps, DeepLIFT, and Occlusion converge on the same clinically relevant region (Figures 5–7). CNNs highlight contiguous ring- or center-focused features, while transformers distribute attention globally. Occlusion aligns most closely with radiologist workflows by revealing how masking regions alters predictions. The consistency across multiple XAI methods reinforces the interpretability and trustworthiness of MammoFormer's decisions.

By pairing each transformer model with optimal preprocessing, applying synthetic augmentation to counter class imbalance, and validating outputs via complementary XAI techniques, MammoFormer bridges transformer potential and clinical reliability. These findings support selecting model–preprocessing pairs tailored to specific architectures, optimizing diagnostic accuracy and ensuring transparent, trustworthy breast cancer detection for real-world screening workflows.

Explainable AI techniques reveal significant architectural differences in how various models process mammographic characteristics. As shown in Figure 5, BaseCNN identifies ring-shaped features distributed throughout the image, while ResNet focuses on central regions with contextual support. DenseTransformer demonstrates a distributed attribution pattern that integrates features across the entire image space, reflecting transformer architecture's inherent capability to capture global relationships through distributed attention points across the spatial domain.

Figure 7 presents targeted attribution analysis where the Occlusion technique specifically highlights regions of highest clinical relevance through bright yellow-orange coloration. These areas represent features that, when occluded, substantially alter model predictions. Multiple XAI techniques consistently identify the same critical features while providing complementary analytical perspectives, with the Occlusion method proving most clinically valuable as its region-based highlighting aligns with radiologists' focus on suspicious areas rather than individual pixels.

The fundamental processing differences between architectures are evident in their visual interpretation patterns. Convolutional neural networks analyze contiguous image regions through local feature extraction, whereas transformer models leverage their capacity for modeling long-range dependencies by distributing attention across the spatial domain. This architectural distinction directly influences how each model type interprets and prioritizes mammographic features.

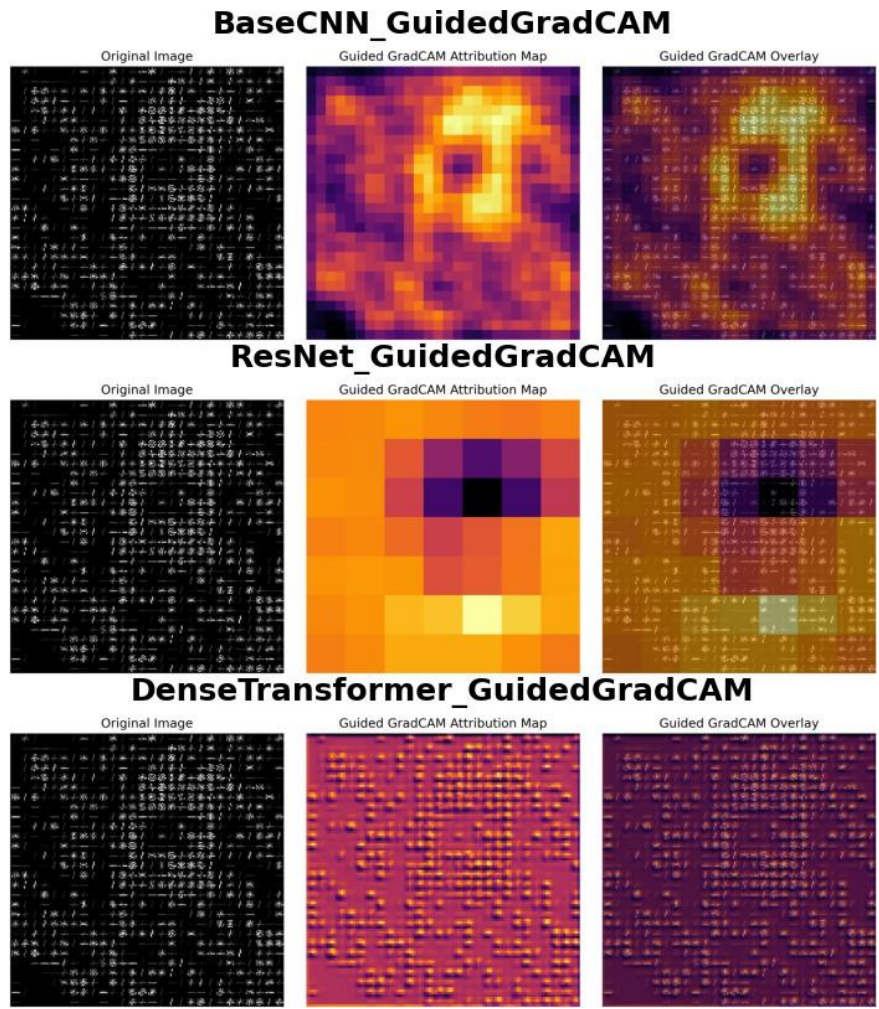

*Figure 5. Guided GradCAM visualization across model architectures: BaseCNN (top) identifies a ring-like structure with strong central activation; ResNet (middle) focuses on a concentrated central region with surrounding contextual features; DenseTransformer (bottom) exhibits distributed attribution patterns throughout the image, demonstrating the transformer model's capacity to integrate information across the entire feature space.*

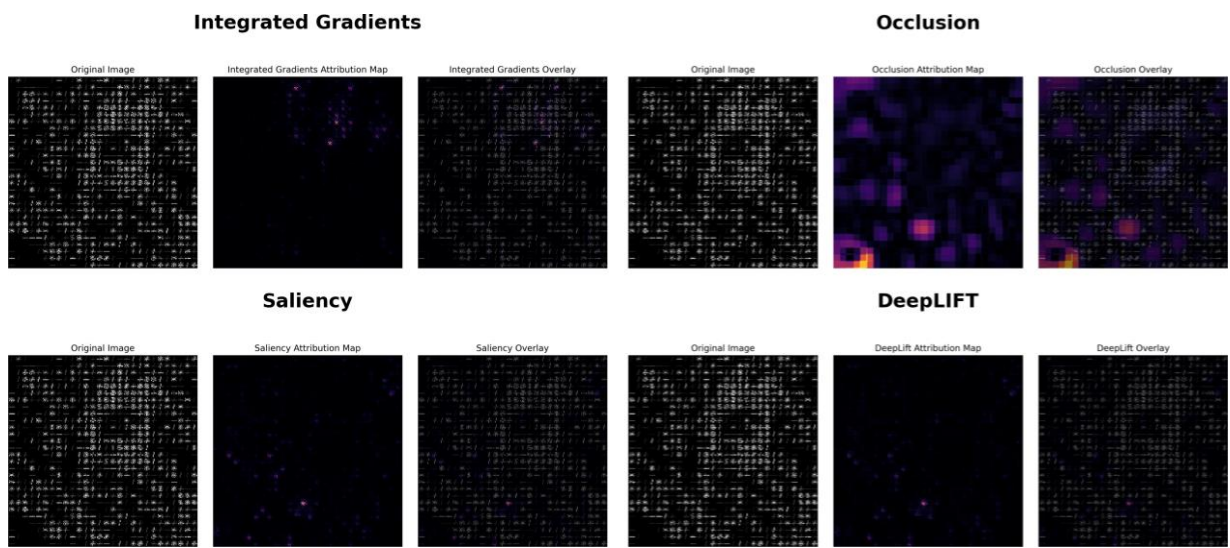

*Figure 6. Comparison of XAI methods on mammogram analysis: Integrated Gradients (top left) highlights sparse point-like features; Occlusion analysis (top right) reveals stronger attribution in lower regions indicating areas that significantly affect prediction when blocked; Saliency maps (bottom left) identify high-gradient features influencing classification; DeepLIFT (bottom right) shows similar sparse attribution patterns to Integrated Gradients with varying intensity.*

The MammoFormer framework demonstrates that transformer-based models, when paired with appropriate feature enhancement techniques, can achieve performance comparable to or exceeding CNNs. HOG features provide the most consistent performance improvements for transformer models, with the framework achieving 98.3% accuracy for Vision Transformer and a 83.3% accuracy increase for Swin Transformer. This systematic evaluation of multiple architectures across various feature enhancement methods addresses significant gaps in existing literature and validates the approach of integrating transformer structures with measurable visual components. The integration of diverse XAI methods enhances confidence in model behavior by providing multiple analytical perspectives on significant features, as evidenced by the consistency across techniques shown in Figure 7. This comprehensive approach to explainable mammographic analysis merges transformer architectures with complete explainability elements, facilitating better integration with clinical workflows and providing valuable guidance for implementing transformer-based approaches in medical imaging applications. Figure 6 demonstrates how different XAI methods interpret the same mammographic image with varying attribution patterns. Occlusion analysis shows the strongest and most localized activation in lower regions, indicating critical diagnostic areas. Integrated Gradients, Saliency, and DeepLIFT exhibit similar sparse, point-like attribution patterns with minimal regional concentration, suggesting different sensitivity levels across explanation techniques.

Figure 7 reveals consistent attribution patterns across all XAI methods focused on a central region of potential abnormality. Occlusion analysis provides the strongest signal with bright yellow-orange highlighting, indicating this area's critical importance to model prediction. Integrated Gradients, Saliency, and DeepLIFT all converge on the same central feature, demonstrating robust consensus across different explanation techniques for identifying clinically relevant areas.

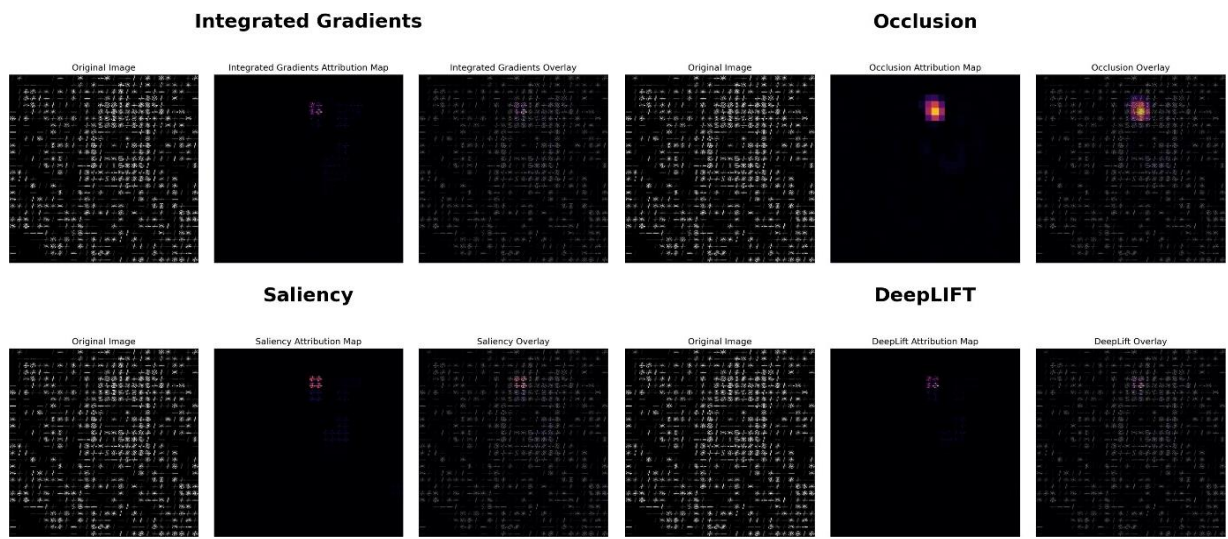

*Figure 7.* XAI analysis of a region containing potential abnormality: Integrated Gradients (top left) shows focused attribution on a central feature; Occlusion analysis (top right) demonstrates strongest activation (bright yellow-orange) in the central region, indicating its critical importance to model prediction; Saliency maps (bottom left) and DeepLIFT (bottom right) both confirm the significance of the same central feature with fine-grained attribution patterns.

Gradients show mainly single points as important features, yet Occlusion analysis presents broader contextual areas to understand model decisions. BaseCNN produces ring-shaped attributions but ResNet focuses its attention on a dense central area and Dense Transformer monitors global image dependencies by engaging in distributed image attention as shown in Figure 5.

The key difference between Figure 6 and Figure 7 is in the presence and clarity of identifiable abnormalities. Figure 6 shows a mammogram with sparse, distributed attribution patterns where XAI methods struggle to identify a clear focal point of interest, resulting in weak and scattered signals across techniques. In contrast, Figure 7 demonstrates strong consensus among all XAI methods, with concentrated attribution patterns focused on a central region containing a potential abnormality. The Occlusion analysis particularly shows dramatic differences - minimal activation in Figure 6 versus bright yellow orange highlighting in Figure 7, indicating the presence of a clinically significant feature that substantially impacts model predictions when occluded.

The synthetic data creation method worked to resolve medical imaging data deficits by producing balance training datasets. The combination of synthetic data generation and multi-feature enhancement strategy that includes original, negative, AHE and HOG provides an effective clinical method to tackle class imbalance successfully.

The Application of several XAI techniques solves the difficulties of" black box" systems by providing transparency for deep learning applications in mammography. Through its clear visual explanations of model decisions MammoFormer assists medical professionals to establish trust because of their ability to understand both model findings and reasoning mechanisms which enables practical healthcare workflow adoption.

*Clinical Implementation Framework*

In real-world practice, MammoFormer is deployed through a phased strategy that aligns with typical clinical workflows while progressively introducing its more advanced capabilities. In the initial phase, Primary Screening, the ResNet-50 classifier serves as the frontline tool for routine mammogram analysis. Integrated seamlessly into existing image-management systems, it processes every incoming study and generates a confidence score alongside GradCAM heatmaps that highlight the most salient regions for each prediction. This familiar visual overlay allows radiologists to review AI-suggested findings within their standard reading environment, leveraging the model's 99.9% accuracy without disrupting established processes.

When a case presents ambiguity, indicated by a lower confidence score or a conflict between the ResNet output and clinical thresholds, the system automatically escalates the study into the Complex Case Analysis phase. Here, the transformer ensemble is activated: the Vision Transformer provides global context reasoning, while the Swin Transformer focuses on architectural distortions. Each model contributes its own explainable-AI visualization, so radiologists can examine both local and global activation patterns side by side. This multi-perspective insight deepens diagnostic reasoning, particularly for subtle lesions or dense-breast contexts where texture and spatial relationships are critical. Finally, MammoFormer full suite, including the primary classifier, transformer validators, and weighted-voting ensemble logic, serves an educational role within radiology training programs. Residents review curated cases under supervision, comparing outputs from each model tier to understand how different architectures interpret the same image. The companion XAI visualizations guide trainees through pattern-recognition exercises, reinforcing the link between algorithmic attention and radiologic signs of malignancy. Over time, this structured exposure builds both confidence in AI-assisted readings and the nuanced judgment required for complex diagnostic decisions.

## 5. Conclusion

MammoFormer bridges the gap between high-performance deep learning and clinically interpretable breast cancer screening by integrating transformer architectures, multi-feature enhancement, and a comprehensive explainability framework. By systematically optimizing transformers with targeted feature engineering, we achieved up to a 13% performance boost, demonstrating that, when properly tuned, these models can rival or exceed traditional CNN accuracy. MammoFormer provides comprehensive guidance for mammography AI implementation through systematic architecture-enhancement optimization. While this study evaluates multiple models to identify optimal combinations, clinical deployment should focus on the most robust solution: ResNet-50 with original images for primary screening (99.9% accuracy). The framework's key contribution lies in demonstrating that transformer models can achieve CNN-comparable performance through appropriate feature enhancement (ViT: 99.0% with HOG, Swin: 96.3% with HOG), and opening opportunities for applications requiring global spatial context modeling. The integrated explainable AI framework provides multi-perspective diagnostic interpretability essential for clinical trust. Future work should focus on clinical validation of these findings and exploration of hybrid approaches combining CNN robustness with transformer global modeling capabilities Our multi-perspective XAI visualizations validate diagnostic reasoning at both local and global levels, fostering radiologist trust and enabling transparent, real-time decision support. The tiered deployment strategy further ensures that MammoFormer can be seamlessly integrated into routine screening for high-confidence cases, escalate to a full ensemble for challenging studies, and serve as an educational tool for radiology trainees. Looking ahead, we will pursue prospective clinical validations and extend our architecture-enhancement principles to other medical imaging modalities, combining mammography with ultrasound and MRI and testing on large, demographically diverse datasets from multiple equipment vendors. By delivering state-of-the-art accuracy alongside workflow-ready explainability, MammoFormer offers a generalizable blueprint for the next generation of AI-driven diagnostic tools, empowering physicians to collaborate with AI systems and ultimately improve patient outcomes.

## Abbreviations

| | |
|---|---|
| AI | Artificial Intelligence |
| ADAS-Cog | Alzheimer's Disease Assessment Scale-Cognitive Subscale |
| AHE | Adaptive Histogram Equalization |
| AUC | Area Under Curve |
| BIRADS | Breast Imaging-Reporting and Data System |
| CBIS-DDSM | Curated Breast Imaging Subset of Digital Database for Screening Mammography |
| CAD | Computer-Aided Detection/Diagnosis |
| CC | Craniocaudal |
| CCT | Compact Convolutional Transformer |
| CNN | Convolutional Neural Network |
| DDSM | Digital Database for Screening Mammography |
| DeepLIFT | Deep Learning Important Features |
| DL | Deep Learning |
| DSC | Dice Similarity Coefficient |
| EDA | Exploratory Data Analysis |
| FLOPS | Floating Point Operations Per Second |
| GradCAM | Gradient-weighted Class Activation Mapping |
| HATNet | Holistic Attention Network |
| HOG | Histogram of Oriented Gradients |
| LSTM | Long Short-Term Memory |
| MIAS | Mammographic Image Analysis Society |
| MIL | Multi-Instance Learning |
| MLO | Mediolateral Oblique |
| MMSE | Mini Mental State Examination |
| MRI | Magnetic Resonance Imaging |
| NLP | Natural Language Processing |
| PET | Positron Emission Tomography |
| PVT | Pyramid Vision Transformer |
| ROC | Receiver Operating Characteristic |
| ROI | Region of Interest |
| SAM | Segment Anything Model |
| TEBLS | Transformer-Encoder-Based Lesion Segmentation |
| ViT | Vision Transformer |
| WSI | Whole Slide Imaging |
| XAI | Explainable Artificial Intelligence |

## Acknowledgments

The authors wish to thank the creators and contributors of the CBIS-DDSM dataset used for this research. Special thanks to our colleagues who provided valuable feedback and technical discussions during the development of the MammoFormer framework. We extend our appreciation to our families for their patience and encouragement throughout this project. Finally, we thank the anonymous reviewers whose insightful comments helped improve the quality of this manuscript.

## Conflicts of Interest

The authors declare no conflicts of interest.

## References

[1] S. Mehta, et al., "End-to-End diagnosis of breast biopsy images with transformers," Medical Image Analysis, vol. 79, 2022. https://doi.org/10.1016/j.media.2022.102466

[2] G. Ayana, et al., "Vision-Transformer-Based Transfer Learning for Mammogram Classification," Diagnostics (Basel), vol. 13, no. 2, 2023. https://doi.org/10.3390/diagnostics13020178

[3] A. A. Jeny, et al., "Hybrid transformer-based model for mammogram classification by integrating prior and current images," Medical Physics, 2025. https://doi.org/10.1002/mp.17650

[4] S. Hussain, et al., "Performance Evaluation of Deep Learning and Transformer Models Using Multimodal Data for Breast Cancer Classification," Cancer Prevention, Detection, and Intervention, Caption 2024, vol. 15199, pp. 59-69, 2025. https://doi.org/10.48550/arXiv.2410.10146

[5] A. Iqbal and M. Sharif, "BTS-ST: Swin transformer network for segmentation and classification of multimodality breast cancer images," Knowledge- Based Systems, vol. 267, 2023. https://doi.org/10.1016/j.knosys.2023.110393

[6] Y. Shen, J. P., F. Yeung, E. Goldberg, L. Heacock, F. Shamout, K. J. Geras, "Leveraging Transformers to Improve Breast Cancer Classification and Risk Assessment with Multi-modal and Longitudinal Data," arxiv.org, 2023. https://doi.org/10.48550/arXiv.2311.03217

[7] X. X. Chen, et al., "Transformers Improve Breast Cancer Diagnosis from Unregistered Multi-View Mammograms," Diagnostics, vol. 12, no. 7, 2022. https://doi.org/10.3390/diagnostics12071549

[8] E. Z. Dalah, et al., "Screening Mammography Diagnostic Reference Level System According to Compressed Breast Thickness: Dubai Health," Journal of Imaging, vol. 10, no. 8, 2024. https://doi.org/10.3390/jimaging10080188

[9] S. Sarker, et al., "MV-Swin-T: MAMMOGRAM CLASSIFICATION WITH MULTI-VIEW SWIN TRANSFORMER," IEEE International Symposium on Biomedical Imaging, ISBI 2024, 2024. https://doi.org/isbi56570.2024.10635578

[10] I. Kassis, et al., "Detection of breast cancer in digital breast tomosynthesis with vision transformers," Scientific Reports, vol. 14, no. 1, 2024. https://doi.org/10.1038/s41598-024-72707-2

[11] W. Lee, et al., "Transformer-based Deep Neural Network for Breast Cancer Classification on Digital Breast Tomosynthesis Images," Radiol Artif Intell, vol. 5, no. 3, p. e220159, 2023. https://doi.org/10.1148/ryai.220159

[12] M. L. Abimouloud, K. B., M. Elleuch, O. Aiadi, and M. Kherallah, "Vision transformer-convolution for breast cancer classification using mammography images: A comparative study," International Journal of Hybrid Intelligent Systems, vol. 20, no. 2, pp. 67-83, 2024. https://doi.org/10.3233/HIS-240002

[13] H. N. Wang, et al., "Transformer-Based Explainable Model for Breast Cancer Lesion Segmentation," Applied Sciences-Basel, vol. 15, no. 3, 2025. https://doi.org/10.3390/app15031295

[14] O. Adeniran, T. Blessing, T. E. Ajibola, O. O. Ejiga Peter, A. O. Ajala, M. M. Rahman, and F. Khalifa, "Explainable MRI-Based Ensemble Learnable Architecture for Alzheimer's Disease Detection," Algorithms, vol. 18, no. 3, pp. 163, 2025. https://doi.org/10.3390/a18030163

[15] W. Wang, et al., "Semi-supervised vision transformer with adaptive token sampling for breast cancer classification," Frontiers in Pharmacology, vol. 13, 2022. https://doi.org/10.3389/fphar.2022.929755

[16] W. S. Lee, et al., "Transformer-based Deep Neural Network for Breast Cancer Classification on Digital Breast Tomosynthesis Images," Radiology-Artificial Intelligence, vol. 5, no. 3, 2023. https://doi.org/10.1148/ryai.220159

[17] A. Basaad, et al., "A BERT-GNN Approach for Metastatic Breast Cancer Prediction Using Histopathology Reports," Diagnostics, vol. 14, no. 13, 2024. https://doi.org/10.3390/diagnostics14131365

[18] O. O. Ejiga Peter, M. M. Rahman, and F. Khalifa, "Advancing AI-Powered Medical Image Synthesis: Insights from MedVQA-GI Challenge Using CLIP, Fine- Tuned Stable Diffusion, and Dream-Booth + LoRA," 2024. https://doi.org/0.48550/arXiv.2502.20667

[19] O. O. Ejiga Peter, "Advancing Colonoscopy Analysis Through Text-to-Image Synthesis Using Generative AI for Intelligent Data Augmentation, Image Classification, and Segmentation," ProQuest, 2024.

[20] O. O. Ejiga Peter, O. T. Adeniran, J. A. MacGregor, F. Khalifa, and M. M. Rahman, "Text-Guided Synthesis in Medical Multimedia Retrieval: A Framework for Enhanced Colonoscopy Image Classification and Segmentation," Algorithms, vol. 18, no. 3, pp. 155, 2025. https://doi.org/10.3390/a18030155

[21] O. O. Ejiga Peter, O. Akingbola, C. Amalahu, O. Adeniran, F. Khalifa, and M. M. Rahman, "Synthetic data-driven multi-architecture framework for automated polyp segmentation through integrated detection and mask generation," Proc. SPIE, vol. 13410, 2025. https://doi.org/10.1117/12.3049369